\documentclass[12pt]{article}

\usepackage{hyperref}
\usepackage{amssymb}
\usepackage{color}

\begin{document}

\renewcommand{\abstractname}{\hfill}

\newpage
\pagenumbering{arabic}

{\Large \bf Some Effects of Different Coordinate Systems

\vspace{4pt}
in Cosmology
}

\vspace{11pt}
{\bf
Andrey A. Grib${}^{1,2}$ and Yuri V. Pavlov${}^{3,4}$
}
\begin{abstract}
${}^{1}$ Theoretical Physics and Astronomy Department, The Herzen  University,
48~Moika, St.\,Petersburg, 191186, Russia; andrei\_grib@mail.ru

${}^{2}$ A. Friedmann Laboratory for Theoretical Physics, St.\,Petersburg, Russia

${}^{3}$ Institute of Problems in Mechanical Engineering of
Russian Academy of Sciences,
61 Bolshoy, V.O., St.\,Petersburg, 199178, Russia; yuri.pavlov@mail.ru

${}^{4}$ N.I.\,Lobachevsky Institute of Mathematics and Mechanics,
Kazan Federal University, Kazan, Russia
\end{abstract}

    \begin{abstract}
\noindent
{\bf Abstract} \
    The analysis of the dynamics of radial movement in different reference
frames used in cosmology is made.
    Use of different frames leads to the difference in inertial forces
resulting in different observable effects.
    The important effect is the appearance in the system different from
the synchronous one of the acceleration proportional to the distance
analogous to the action of the cosmological constant.
    Numerical estimate of the difference of this effective cosmological
constant and the invariant constant in Einstein equations is made.

\vspace{4pt}
\end{abstract}

\section{\normalsize Introduction}
\label{secIn}

    In classical mechanics one has inertial reference frames as
preferred ones.
    The description of motion in such frames usually is simpler than in
the case of use of noninertial frames.
    In the last case one must introduce the so called inertial forces
for example the centrifugal force and the Coriolis force in case of
the rotating  reference frame.
    However sometimes it is not reasonable to use the inertial frame.
    For example noninertial  reference frame formed by rotating Earth is
natural if one wants to describe all observations on the immovable Earth.

    In General Relativity there are no preferred coordinate systems.
    If two different coordinate systems describe the same space-time region
they equally can be used.
    However as it is the case for classical mechanics it is necessary to
understand the physical sense of different terms in equations of movement of
bodies in different coordinate systems.
    In cosmology the well known example of dipole and quadrupole terms in
background radiation can be mentioned~\cite{Kogut,NaselskiiNovikovN}.
    Our telescopes on the Earth or in space of the Solar system correspond not
to the synchronous reference frame in cosmology usually used in it.
    So the problem of the interpretation of appearance of terms arising due
to the reference frame used in physical cosmology is arising.

    The difference between reference frames is important for large distances
where the nondiagonal term in metrical tensor is not small.
    It is well known for the case of rotating reference frame that the Doppler
effect and the redshift of light in it is different from the inertial
case~\cite{Kundig63}.
    Measuring the redshift is the main method to get information on
far Galaxies and their movement.
    Information on dark energy and cosmological constant is obtained
mainly from these measurements.

    And it is well known from the text book~\cite{LL_II}, Sec.~88,
Problem one, that
the forces acting on a particle in a gravitational field are obtained
from the analysis of the Christoffel terms in geodesic equations.
    That is why we must write these equations and expressions for such
terms to understand the origin of noninertial forces appearing in our case.

     In this paper we write geodesic equations in the homogeneous expanding
Universe in two different reference frames often used in cosmology and
investigate the physical meaning of different terms arising in them.
    The expressions for the energy and momentum are also obtained.
    In our previous papers~\cite{GribPavlov2016d,GribPavlov2019h,GribPavlov2020}
we showed that besides the difference in inertial forces there is a difference
in possible energies of particles, i.e. existence of particles with negative
energies in the nonsynchronous system.
    The analysis of geodesic equations made by us shows that besides
the usual existence of some kind of the viscosity force existing in
the synchronous frame leading to deceleration there appears a term leading
in some cases to the acceleration.
    This acceleration occurs however for the cases of nonzero cosmological
constant or exotic matter equation of state which seems to be
the expectable result.

\section{\normalsize Isotropic homogeneous Universe and Einstein's equation}
\label{secNEEU}

    The square of space-time interval of the isotropic cosmological Friedmann model
can be written in synchronous frame~\cite{LL_II} as
    \begin{equation}
d s^2 = c^2 d t^2 - a^2(t) \left( \frac{d r^2}{1 - K r^2} + r^2 d \Omega^2 \right),
\label{f1}
\end{equation}
    where $c$ is speed of light,
parameter $r$ is changing from 0 to $\infty $ in open ($ K= -1 $)
and quasi-Euclidean flat models ($ K=0 $),
and from  0 to 1 in closed cosmological model ($ K= 1 $),
$d \Omega^2 = d \theta^2 + \sin^2 \theta \, d \varphi^2 $.
    Changing the variables
    \begin{equation}
r = f(\chi) = \left\{
\begin{array}{ll}
\sin{\chi} , & \ \ \ K=1, \\
\chi , & \ \ \ K=0, \\
\sinh{\chi} , &\ \ \ K=-1
\end{array}
\right.
\label{zamp}
\end{equation}
    the metric~(\ref{f1}) can be written also in form
    \begin{equation}
d s^2 = c^2 d t^2 - a^2(t)
\left( d \chi^2 + f^2(\chi) d \Omega^2 \right).
\label{f1n}
\end{equation}
    In closed model $\chi$ is changing varying from 0 to $\pi$,
in cases $K=0,-1$ one has $\chi \in [0, + \infty)$.
    Using the conformal time~$\eta$:
    \begin{equation}
c d t = a(\eta) \, d \eta,
\label{eta}
\end{equation}
    the metric~(\ref{f1n}) takes the form
    \begin{equation}
d s^2 = a^2(\eta) \left( d \eta^2 - d \chi^2 - f^2(\chi) d \Omega^2 \right).
\label{etaf1}
\end{equation}

    The Christoffel symbols
    \begin{equation}
\Gamma^{\, i}_{\, kl} = \frac{1}{2} g^{im} \left(
\frac{\partial g_{mk}}{\partial x^l} + \frac{\partial g_{ml}}{\partial x^k}
- \frac{\partial g_{kl}}{\partial x^m} \right)
\label{Gijk}
\end{equation}
    in the homogeneous isotropic space-time with metric~(\ref{etaf1}) are
     \begin{equation}
\Gamma^{\,0}_{\, 00}= \frac{a'}{a} = \frac{\dot{a}}{c}   \ ,
\ \ \ \  \Gamma^{\,i}_{\, 0 j}=\frac{\dot{a}}{c} \, \delta^i_j   \ ,
\ \ \ \ \Gamma^{\,0}_{\, \alpha \beta}= \frac{\dot{a}}{c} \, \gamma_{\alpha \beta} \ ,
\ \ \ \ \Gamma^{\,\alpha}_{\, \beta \delta}(g_{ik}) =
\Gamma^{\,\alpha}_{\, \beta \delta}(\gamma_{\nu \mu})
\label{GGG}
\end{equation}
    where prime denotes the derivative with respect to conformal time~$\eta$,
the dot above symbol is the derivative with respect to time~$t$,
$ \gamma_{\alpha \beta} $ is metric of 3-dimensional space of constant curvature~$K$.

    Ricci tensor components and the scalar curvature are
    \begin{equation}
R_{00} = 3 \frac{a \ddot{a}}{c^2}, \ \ \ R_{\alpha \beta } =
- \gamma_{\alpha \beta } \left[ \frac{a \ddot{a}}{c^2} +
2 \left( \frac{\dot{a}^2}{c^2} + K \right) \right],
\label{Rik}
\end{equation}
     \begin{equation}
R = \frac{6}{a^2} \left[ \frac{a \ddot{a}}{c^2} +
\left( \frac{\dot{a}^2}{c^2} + K \right) \right].
\label{Rska}
\end{equation}
    Einstein's equations are
     \begin{equation}
R_{ik} -\frac{1}{2} R g_{ik} + \Lambda g_{ik} = - 8 \pi \frac{G}{c^4} T_{ik},
\label{Eur}
\end{equation}
    where $\Lambda$ is cosmological constant,
$G$ is the gravitational constant,
$T_{ik}$ is energy-momentum tensor of background matter.
    Equations~(\ref{Eur}) for metric~(\ref{etaf1}) are
     \begin{equation}
\frac{\dot{a}^2 + K c^2}{a^2} = \frac{ c^2}{ 3 }
\left( \frac{ 8 \pi G }{ c^4 } T_0^0 + \Lambda \right),
\label{Eur0}
\end{equation}
     \begin{equation}
\frac{ \ddot{a} }{a} + \frac{\dot{a}^2 + K c^2}{2 a^2}
= \frac{ c^2}{ 2 } \left( \frac{ 8 \pi G }{ 3 c^4 }
\sum_{\alpha = 1}^3 T_\alpha^\alpha + \Lambda \right).
\label{Eur1}
\end{equation}
    From~(\ref{Eur0}), (\ref{Eur1}) one obtains
     \begin{equation}
\frac{ \ddot{a} }{a} = \frac{ c^2}{ 3 } \left( \Lambda + \frac{4 \pi G }{ c^4}
\left( \sum_{\alpha = 1}^3 T_\alpha^\alpha - T_0^0 \right) \right).
\label{Ehh4}
\end{equation}
    In comoving coordinates the energy-momentum tensor of background matter
in isotropic homogeneous Universe is diagonal
     \begin{equation}
T_i^k = {\rm diag}\, (\varepsilon, -p, -p, -p),
\label{Tikd}
\end{equation}
    where $\varepsilon $ and $p$ are the energy density and pressure of
background matter.
    So
    \begin{equation}
\frac{ \ddot{a} }{a} = \dot{H} + H^2 = \frac{\Lambda c^2}{3} -
\frac{ 4 \pi G } { 3 c^2} \left( \varepsilon + 3 p \right),
\label{ELhh4}
\end{equation}
    where $H= \dot{a}/a$ is the Hubble ``constant''.
    We call it the ``constant'' following~\cite{MTW} in spite it
is a variable depending on time.

    The radial distance between points $\chi=0$ and $\chi$
in metrics~(\ref{f1n}), (\ref{etaf1}) is $D= a(t) \chi$
and it's the same in the metric~(\ref{f1}).
    If $t$ is fixed  then the maximal value of $D$ in the closed model
is $D_{\rm max} =\pi a(t)$.
    In open and flat models $D$ is non limited.
    Nonzero $D$ can be understood as the distance to the far Galaxy from
the place where the observer is located $D=0$.
    The distance~$D$ corresponds to the ``proper distance'' of
book~\cite{Weinberg} and is equal to the distance that would be measured by
observers located rather closely in the expanding universe
between the origin of coordinate system and the point with comoving
coordinate $\chi$ at the same moment of cosmic time $t$.

    Take the new coordinates $t, D, \theta, \varphi$
(see also~\cite{Ellis93,Grib95}).
    Then one obtains
    \begin{equation}
d D = \frac{\dot{a}}{a} D\, d t + a \, d \chi , \ \ \ \
d \chi = \frac{1}{a } \left( d D - \frac{\dot{a}}{a}\, D d t \right)
\label{f3}
\end{equation}
    and the interval~(\ref{f1n}) becomes
    \begin{equation}
d s^2 = \left( 1 - \left( \frac{ H D }{c } \right)^2 \right) c^2 d t^2
+ 2 H D\, d D d t - d D^2 - a^2 f^2(D/a)\, d \Omega^2 .
\label{f4}
\end{equation}
    Note that metric~(\ref{f4}) is not singular on the surface $D=c/H$,
in spite of  $g_{00} =0$.
    The ${\rm det} \left( g_{ik} \right) = - a^4 f^4(D/a)\sin^2 \theta $ is zero
only for $D=0$ or $\theta =0,\pi$, where the coordinate singularities are conditioned
by use of spherical coordinates of 3-space.
    The surface $D=c/H$ is analogous to the static limit for
rotating black hole (see~\cite{GribPavlov2019h}).
    It is also called apparent horizon in cosmology~\cite{Faraoni11}.

    The energy-momentum tensor for background matter in coordinates $t, D$
can be found from formula
     \begin{equation}
T_{ik} =(p+\varepsilon )u_i u_k - p g_{ik}
\label{TikdG}
\end{equation}
    (see~\cite{LL_II}) and expressions for four-velocities $u^i$
of the background matter (with $\chi={\rm const}$)
     \begin{equation}
u^i = \left( 1, \frac{HD}{c}, 0, 0 \right), \ \ \ \
u_i = \left( 1, 0, 0, 0 \right).
\label{uik}
\end{equation}
    That is why one has
     \begin{equation}
T_{0}^0 =\varepsilon , \ \ \ \ T_\alpha^\beta = -p \delta_\alpha^\beta
\label{TikdD}
\end{equation}
    and therefore
$\varepsilon $ and $p$ are the energy density and pressure of
background matter in coordinates $t,D$ also.

    The same result can be obtained by coordinate transformation of
the tensor~(\ref{Tikd}) from one coordinate system to another.
    The difference in the form of the energy-momentum tensor in
coordinates $t$, $D$ is manifested in the appearance of non-diagonal terms.

    Note that for small distances ($ D H /c \ll 1 $, $D/a \ll 1$)
the metric~(\ref{f4}) becomes the metric of comoving spherical coordinate system
    \begin{equation}
d s^2 = c^2 d t^2 - d D^2 - D^2  d \Omega^2 .
\label{f4D0}
\end{equation}
    If the object is located far from the observer
($ D H /c \ll 1 $ is not correct) one cannot go to the diagonal form~(\ref{f4D0}).

\section{\normalsize Free movement in different coordinate systems}
\label{secKerrVr}

    Let us write the equation for the geodesic lines
    \begin{equation}
\frac{d^2 x^i}{d \lambda^2} = - \Gamma^{\, i}_{kl} \frac{d x^k}{d \lambda}
\frac{d x^l}{d \lambda},
\label{GenGeod}
\end{equation}
    where $\lambda$ is the affine parameter on the geodesic.
    The terms in the right hand side are similar to inertial forces
in nonrelativistic classical mechanics.

    First study the radial movement in coordinates $\eta, \chi$.
    Equations of radial geodesics in these coordinates taking
into account~(\ref{GGG}) are
    \begin{equation}
\frac{d^2 \eta}{d \lambda^2} + \frac{\dot{a}}{c}
\left[  \left( \frac{d \eta}{d \lambda} \right)^2 +
\left( \frac{d \chi}{d \lambda} \right)^2 \right] = 0,
\label{Geodeta0}
\end{equation}
    \begin{equation}
\frac{d^2 \chi}{d \lambda^2} + 2 \frac{\dot{a}}{c}
\frac{d \eta}{d \lambda} \frac{d \chi}{d \lambda} = 0 .
\label{Geodeta1}
\end{equation}
    In order to give the interpretation of dissipative terms
in~(\ref{Geodeta1}) let us find the energy and momentum of the particle in
the corresponding coordinate system.
    Note that the geodesic equations can be obtained from
the Lagrangian~\cite{Chandrasekhar}
    \begin{equation}
L = \frac{ g_{ik} }{2} \, \frac{ d x^i}{d \lambda} \frac{ d x^k}{d \lambda}.
\label{Lgik}
\end{equation}
    Generalized momenta are by definition
    \begin{equation}
p_i  \stackrel{\rm def}{=} \frac{\partial L}{\partial \dot{x}^i}
= g_{ik} \frac{d x^k}{d \lambda } ,
\label{Lpdef}
\end{equation}
    where now $ \dot{x}^i = d x^i/d \lambda $.
    The value of $p_i p^i $ is conserved due to Euler-Lagrange equations
    \begin{equation}
\frac{d }{d \lambda} \frac{\partial L}{\partial \dot{x}^n} -
\frac{\partial L}{\partial x^n} = 0.
\label{LEL}
\end{equation}
    For time-like geodesics the affine parameter can be taken as
$ \lambda = \tau / m$ so that $\tau$ is the proper time of the particle
and then
    \begin{equation}
p_i p_k g^{ik} = m^2 c^2.
\label{pikmc}
\end{equation}

    If the metric components $g_{ik}$ do not depend on some coordinate $x^n$
then the corresponding canonical momentum (the corresponding covariant component)
is conserved in motion along the geodesic due to Euler-Lagrange equations:
    \begin{equation}
\frac{\partial g_{ik}}{\partial x^n} = 0 \ \ \Rightarrow \ \
p_n = \frac{\partial L}{\partial \dot{x}^n} = g_{nl}
\frac{d x^l}{d \lambda} = {\rm const}.
\label{okuel}
\end{equation}
    Note that contravariant components of the 4-momentum correspond to
the 4-velocities multiplied by the particle mass
$p^n = m d x^n / d \tau $ and are generally not conserved
even in case when metric does not depend in the corresponding coordinates.

    The covariant radial component of the momentum of the particle in
the homogeneous isotropic expanding space with metrics~(\ref{f1n})
and~(\ref{etaf1}) are
    \begin{equation}
p_\chi = - a^2(t) \frac{d \chi}{d \lambda} =
- m a^2(t) \frac{d \chi}{d \tau} .
\label{imp}
\end{equation}
    The covariant radial component of the momentum for metric~(\ref{f4}) is
    \begin{equation}
p_D = - \frac{d D}{d \lambda} + H D \frac{d t}{d \lambda} = \frac{p_\chi}{a}.
\label{impD}
\end{equation}
    One can see from~(\ref{imp}), (\ref{impD}) that $ p_D $ depends on
the velocity $d \chi / d \tau$ but not on the value of~$\chi$.
    Contravariant radial component for metric~(\ref{f4}) is
    \begin{equation}
p^D = m \frac{d D}{d \tau} = ma \frac{d \chi}{d \tau} + m D H\frac{d t}{d \tau}.
\label{impDK}
\end{equation}
    For $D \to 0$ the metric~(\ref{f4}) becomes the metric of comoving
coordinate system~(\ref{f4D0}),
so let us call $ \tilde{p}^D = ma\, d \chi / d \tau = - p_\chi/ a$
the ``physical'' radial component of the momentum of the particle.

    For radial movement $p_\chi = {\rm const}$,
because the metric components $g_{00}$, $g_{01}$, $g_{11}$
don't depend on $\chi$ (see~(\ref{etaf1}) and~(\ref{okuel})).
    So after expanding of space in $k$-times,
the ``physical'' momentum $ - p_\chi/ a$,  becomes smaller in k-times.

    The energy defined by translations in time~$\eta$ is equal to
    \begin{equation}
E_\eta = p_\eta c = m c a^2 \frac{d \eta}{d \tau}.
\label{EEt}
\end{equation}
    For metric~(\ref{f4}) due to~(\ref{Lpdef}) one obtains
    \begin{equation}
E_{D} = p_t c = mc^2 \left( 1 - \left( \frac{ H D }{c } \right)^2 \right)
\frac{d t}{d \tau} + m H D \frac{d D}{d \tau}.
\label{EEtf4}
\end{equation}
    Let us call the ``physical'' energy $E$ the energy measured by the observer
in the reference frame of the background matter in which at the given moment
the particle is at the origin.
    Going to the limit $D \to 0$ in~(\ref{EEtf4}) we obtain
    \begin{equation}
E = mc^2 \frac{d t}{d \tau} = \frac{E_\eta}{a}.
\label{Ef}
\end{equation}
    Note that $E$ is equal to $E_D$ only for $D \to 0$.

    From~(\ref{EEt}) and (\ref{Ef}) equation~(\ref{Geodeta1})
can be written as
    \begin{equation}
\frac{d^2 \chi}{d \tau^2} + 2 H \frac{E}{mc^2} \frac{d \chi}{d \tau} = 0 .
\label{Geod1et}
\end{equation}
    So radial movement in expanding Universe for the observer with
coordinate~$ \chi $ is similar to movement in viscous medium with viscosity
proportional to the Hubble constant.

    Writing~(\ref{Geod1et}) in the form
    \begin{equation}
m a \frac{d^2 \chi}{ d \tau^2} = - \frac{2 E}{mc^2} H \tilde{p}^D ,
\label{impd}
\end{equation}
    one obtains that the ``inertial'' force for the coordinate $ \chi$ is
equal to the doubled specific energy of the body with the minus sign multiplied
on the ratio of the ``physical'' momentum of the particle to the Hubble
time $t_H= 1/H$.

    From equations~(\ref{Geodeta0}), (\ref{Geodeta1}) we find, that
    \begin{equation}
\frac{d^2 \chi}{ d t^2} + H \frac{d \chi}{ d t} \left( 4 +
\left( \frac{a d \chi}{c d t} \right)^2 \right)= 0 .
\label{GeoDt}
\end{equation}
    For case of non-relativistic movement relative to background matter
one has $|a d \chi / c d t | \ll 1$ and therefore
    \begin{equation}
m a \frac{d^2 \chi}{ d t^2} \approx -  4 m a H \frac{d \chi}{ d t}
= - 4 H \tilde{p}^D .
\label{GeoDtn}
\end{equation}
    So for the observer using not the proper time of the moving particle but
the coordinate time $t$, the ``inertial'' force is equal to the ratio of
the ``physical'' momentum of the particle to the Hubble time $t_H$
multiplied by minus four.

    One can find the dependence of the radial component in time in case of
radial movement of point mass~$m$ with fixed conserved radial component
of momentum~$-p_\chi$ at point~$\chi_0$ in time~$t_0$ from
the equations~(\ref{pikmc}), (\ref{imp})
    \begin{equation}
\chi (t)  =\chi_0 + \frac{p_\chi}{m} \int \limits_{t_0}^t
\frac{d t}{\displaystyle a^2 \sqrt{ 1 + \left( \frac{p_\chi}{m c a} \right)^2 } }.
\label{urd}
\end{equation}

    Now let us consider radial movement in coordinates $(t, D)$.
    Changing the variables in~(\ref{Geodeta0}), (\ref{Geodeta1}),
one obtains the radial geodesic equations in coordinates $(t, D)$
    \begin{equation}
\frac{d^2 t}{d \lambda^2} + \frac{H}{c^2}
\left( \frac{d D}{d \lambda} - D H \frac{d t}{ d \lambda} \right)^2 = 0,
\label{Geod0}
\end{equation}
    \begin{equation}
\frac{d^2 D}{d \lambda^2} + \frac{D H^2}{c^2}
\left( \frac{d D}{d \lambda} - D H \frac{d t}{ d \lambda} \right)^2
-(\dot{H} + H^2) D \left(\frac{d t}{ d \lambda} \right)^2 = 0.
\label{Geod1}
\end{equation}
    These equations can be obtained directly from~(\ref{GenGeod})
using the Christoffel symbols in metric~(\ref{f4})
     \begin{eqnarray}
&&\Gamma^{\,0}_{\, 00}= \frac{D^2 H^3}{c^3}  \ ,
\ \ \ \  \Gamma^{\,0}_{\, 0 1}= - \frac{ D H^2 }{c^2}  \ ,
\ \ \ \  \Gamma^{\,0}_{\, 1 1}= \frac{H}{c} ,
\nonumber \\
&& \Gamma^{\,1}_{\, 0 0}= \frac{ D}{c^2}
\left( \frac{D^2 H^4}{c^2} - \dot{H} - H^2 \right),
\ \ \ \ \Gamma^{\,1}_{\, 01}= - \frac{D^2 H^3}{c^3}  \ ,
\ \ \ \  \Gamma^{\,1}_{\, 1 1}= \frac{ D H^2 }{c^2} .
\label{GGGnk}
\end{eqnarray}
    Note that due to~\cite{LL_II}, Sec.~87 terms
$m \Gamma^{\,i}_{\, kl} \frac{d x^k}{d \lambda} \frac{d x^l}{d \lambda}$
play the role of forces acting on the particle with mass~$m$
in gravitational field.
    It is evident that such forces depend on the choice of the reference
frame.
    In absence of gravitation due to the possibility of inertial reference
frame one can discriminate inertial forces from other forces.
    In general case such differentiation is impossible.

    One of the solution of the system of equation~(\ref{Geod0}), (\ref{Geod1}) is
    \begin{equation}
D = \frac{a}{a_0} D_0, \ \ \ \frac{d t}{d \lambda } = {\rm const}.
\label{Resh1}
\end{equation}
    This solution describes the world line of the particle at rest in
synchronous with the background matter reference frame~(\ref{f1n}):
$\chi= \chi_0$
and proper time is equal to the coordinate time~$t$ (see also~\cite{Faraoni21}).

    The equation of movement~(\ref{Geod1}) by using~(\ref{impD}), (\ref{Ef})
can be written as
    \begin{equation}
\frac{d^2 D}{d \tau^2} = - D H^2 \left( \frac{p_D}{mc} \right)^2
+ (\dot{H} + H^2) D \left(\frac{E}{ mc^2} \right)^2.
\label{Geod1D}
\end{equation}
    In nonrelativistic case the first term proportional to the
square of the velocity corresponds to the force of resistance for movement
in medium.
    The force of this resistance is proportional to the square of the
Hubble constant and the distance from the point of observation.
    In nonrelativistic case $p_D \approx m v$, where $v$ is the velocity
of the moving body relative to background matter.
    So the first term is in $(v/c)^2$ times smaller than the second.

    The second term in nonrelativistic case ($E \approx mc^2$) does not
depend on the velocity and is proportional to the distance~$D$ from
the coordinate origin, i.e. corresponds to some cosmological constant.

    Consider this issue more exactly.
    Let us write the interval for the case of ``point'' mass~$M$ in presence
of cosmological constant~\cite{Tolman} (Kottler metric~\cite{Kottler18})
    \begin{equation}
ds^2 = \left( 1 - \frac{2 G M}{c^2 r} - \frac{\Lambda r^2}{3} \right) c^2 dt^2
- \frac{dr^2}{\displaystyle 1 - \frac{2 G M}{c^2 r} - \frac{\Lambda r^2}{3} }
- r^2 ( d \theta^2 + \sin^2 \theta d \phi^2 ).
\label{Kottler}
\end{equation}
    In nonrelativistic case (see Sec.~99 in~\cite{LL_II}) this metric
leads to gravitational potential
    \begin{equation}
\varphi = - \frac{GM}{r} - \frac{ \Lambda c^2 r^2}{6}
\label{Kotphi}
\end{equation}
    and to the acceleration of test body
    \begin{equation}
\frac{d^2 r}{d t^2} = - \frac{GM}{r^2} + \frac{ \Lambda c^2 r}{3} .
\label{Kota}
\end{equation}
    So in nonrelativistic case the cosmological constant describes the force
proportional to the distance to the observable body~\cite{McVittie}.

    From~(\ref{Geod0}), (\ref{Geod1}) one can obtain the equation of
radial movement of the test body in homogeneous isotropic
cosmology in coordinates $t, D$, written as
    \begin{equation}
\frac{d^2 D}{d t^2} = \frac{H}{c^2} \left( \frac{d D}{d t} - D H \right)^3
+ (\dot{H} + H^2) D .
\label{Geod1Dt}
\end{equation}
    Due to $dD / dt - D H = a d \chi / d t$ the first term describes
in analogy with the mechanics of continuous media
the drugging of the body by the moving viscous medium if the body is
moving in the same direction as medium and it describes deceleration
if the body moves in the opposite direction.
    The arising inertial force is proportional to the cube of the relative
velocity.

    The second term as in case~(\ref{Geod1D}) is similar to action of
the cosmological constant.
    Comparing this term with~$\Lambda$ in equation~(\ref{Kota}),
one obtains the ``effective'' cosmological constant $\Lambda_{\rm eff}$
    \begin{equation}
\Lambda_{\rm eff} = \frac{3}{c^2} (\dot{H} + H^2) .
\label{LamDt}
\end{equation}
    Due to~(\ref{ELhh4}) one obtains the relation of this effective cosmological
constant with the constant present in Einstein's equation and the energy
density and pressure in homogeneous isotropic cosmological model
    \begin{equation}
\Lambda_{\rm eff} =
\Lambda - \frac{ 4 \pi G } { c^4} \left( \varepsilon + 3 p \right).
\label{Lameff}
\end{equation}
    One can use~(\ref{ELhh4}) for evaluation~(\ref{Lameff}) in
the comoving system because~$H$ is the same in our different frames.

    So nonaccurate use in cosmology of coordinates $t,D$
can lead to the measurement of $ \Lambda_{\rm eff} $ instead of true
cosmological constant~$ \Lambda $.

    Let us evaluate $ \Lambda - \Lambda_{\rm eff}$ for real Universe
taking $p=0$ and the energy density of visible and dark matter
$\approx31$\% of the critical density
$\varepsilon_c = 3 H^2 c^2/(8 \pi G)$~\cite{Planck}
so that $\Lambda$ term (or dark energy) is approximately $\approx69$\%:
    \begin{equation}
\Lambda - \Lambda_{\rm eff} =
\frac{ 4 \pi G } { c^4} \left( \varepsilon + 3 p \right) \approx
0.31 \frac{3 H^2}{2 c^2} \approx 0.22 \Lambda.
\label{DLamef}
\end{equation}
    So use of $t,D$ coordinates and usual nonrelativistic
expression~(\ref{Kota}) for cosmological constant can lead to the error
of the order of 20\%.

    If $ \Lambda =0 $ one can see from~(\ref{Lameff})
that in order to have the true sign ``+'' for
the $\Lambda_{\rm eff}$ as in the case of observable cosmological
constant~\cite{Schmidt98}--\cite{Perlmutter99} it is necessary to suppose
the dominance of exotic background matter
(quintessence etc.) with $ \varepsilon + 3 p<0 $.

\section{\normalsize Conclusion}
\label{secConcl}

    Our analysis shows that for any measurement
of cosmological distances it is necessary to use the dynamical equations in
corresponding coordinates to exclude the effects similar to inertial forces.
    In this paper we show that these forces can lead not only to deceleration
of galaxies but also to acceleration.
    However this acceleration cannot mean that inertial force plays
 the role of ``dark energy'' because
it occurs only in case of positive cosmological constant which usually is
considered as ``dark energy''.
    Nevertheless our analysis shows that due to the effect of the used
coordinate system the measured value of the cosmological
constant $\Lambda_{\rm eff}$ can be different from $ \Lambda $ ---
the cosmological constant in Einstein's equations.
    It is the matter of fact that our above estimate of 20\% varies during
the evolution of the Universe.
    Really, if the Universe would be described by the de Sitter metric, then,
taking $\varepsilon = p =0$, one easily obtains $ \Lambda_{\rm eff} = \Lambda $.
    The same is valid for the Kottler metric.
    In these cases there is no need in ``dark energy''
with $ \varepsilon + 3 p<0 $.
    For the Universe described by the metric of the Friedmann stage going to
the de Sitter stage the effective slowly varying cosmological constant
converges to the value of the invariant cosmological constant.
    The effective cosmological constant is non invariant being calculated
through the Christoffel symbol which is not a tensor.
    It is proportional to the square of the Hubble constant depending on time
and becomes really constant at the De Sitter stage.
    But then one sees that the invariant constant, which has the same value,
also depends on the square of the Hubble constant at some fixed moment of time
characterizing the transfer to the De Sitter stage.
    It cannot have  arbitrary value.

    One of the much disputed matters concerning
the observable value of the cosmological constant is its proportionally
to $H_0^2$ value, where $H_0$ is the modern value of Hubble constant
(anthropic principle etc.).
    Strange as it is but term of the same order as we have shown in our paper
appears due to use of $t, D$ coordinate system.
    The message of our paper is not the doubt in existence of nonzero
cosmological constant but in careful analysis of what we really observe
using our instrumentation and what coordinate system corresponds to it.

\vspace{9pt}
\noindent
{\bf Acknowledgements} \
    The work of Yu.V.P. was supported by the Russian Government Program of
Competitive Growth of Kazan Federal University.


\end{document}